\pdfoutput=1

\documentclass[11pt,twoside,a4paper,cmspaper,final,collab]{cms-tdr}
\def\svnVersion{493567P}\def\cmsCernNoTag{CERN-EP-2018-328}\def\cmsCernDate{\today}\def\cmsMessage{Published in Physical Review Letters as \href{http://dx.doi.org/10.1103/PhysRevLett.122/132003}{\doi{10.1103/PhysRevLett.122/132003.}}}

\begin{document}\cmsNoteHeader{TOP-18-008}

\hyphenation{had-ron-i-za-tion}
\hyphenation{cal-or-i-me-ter}
\hyphenation{de-vices}
\RCS$HeadURL: svn+ssh://svn.cern.ch/reps/tdr2/papers/TOP-18-008/trunk/TOP-18-008.tex $
\RCS$Id: TOP-18-008.tex 493554 2019-04-08 08:41:28Z wverbeke $

\ifthenelse{\boolean{cms@external}}
{\providecommand{\suppMaterial}{the supplemental material~\cite{suppMat}}}
{\providecommand{\suppMaterial}{Appendix~\ref{app:suppMat}}}
\providecommand{\cmsTable}[1]{\resizebox{\textwidth}{!}{#1}}

\cmsNoteHeader{TOP-18-008}
\title{Observation of single top quark production in association with a \texorpdfstring{\PZ boson in proton-proton collisions at $\sqrt{s} = 13\TeV$}{Z boson in proton-proton collisions at sqrt(s) = 13 TeV}}

\newcommand{\tZq}{\cPqt\cPZ\cPq\xspace}
\newcommand{\ttZ}{\ensuremath{\cPqt\cPaqt\cPZ}\xspace}
\newcommand{\ttW}{\ensuremath{\cPqt\cPaqt\PW}\xspace}
\newcommand{\tth}{\ensuremath{\cPqt\cPaqt\PH}\xspace}
\newcommand{\ttgamma}{\ensuremath{\cPqt\cPaqt\gamma^{(*)}}\xspace}
\newcommand{\tWZ}{\cPqt\PW\cPZ\xspace}
\newcommand{\ttVV}{\ensuremath{\ttbar\cmsSymbolFace{VV}}\xspace}
\newcommand{\ttV}{\ensuremath{\ttbar\cmsSymbolFace{V}}\xspace}
\newcommand{\WZ}{\ensuremath{\PW\cPZ}\xspace}
\newcommand{\ZZ}{\ensuremath{\cPZ\cPZ}\xspace}
\newcommand{\Zgamma}{\ensuremath{\cPZ\gamma^{(*)}}\xspace}
\newcommand{\XgammaStar}{\ensuremath{\cmsSymbolFace{X}\gamma^{(*)}}\xspace}
\newcommand{\abseta}{\ensuremath{\abs{\eta}}\xspace}

\date{\today}

\abstract{
The observation of single top quark production in association with a \PZ boson and a quark (\tZq) is reported. Events from proton-proton collisions at a center-of-mass energy of 13\TeV containing three charged leptons (either electrons or muons) and at least two jets are analyzed. The data were collected with the CMS detector in 2016 and 2017, and correspond to an integrated luminosity of 77.4\fbinv. The increased integrated luminosity, a multivariate lepton identification, and a redesigned analysis strategy improve significantly the sensitivity of the analysis compared to previous searches for \tZq production. The \tZq signal is observed with a significance well over five standard deviations. The measured \tZq production cross section is $\sigma (\Pp\Pp \to \tZq \to \cPqt \ell^{+} \ell^{-} \cPq )=111 \pm13 \stat \, _{-9}^{+11} \syst \unit{fb}$, for dilepton invariant masses above 30\GeV, in agreement with the standard model expectation.}

\hypersetup{
pdfauthor={CMS Collaboration},%
pdftitle={Observation of single top quark production in association with a Z boson in proton-proton collisions at sqrt(s) = 13 TeV},%
pdfsubject={CMS},
pdfkeywords={CMS, physics, top quark}}

\maketitle

The CERN LHC has delivered proton-proton ($\Pp\Pp$) collisions with an unprecedented luminosity at a center-of-mass energy of 13\TeV over the last few years. The large number of high-energy collisions recorded to date allows the probing of very rare standard model (SM) processes. One such process is electroweak (EW) production of a single top quark in association with a $\cPZ$ boson and a quark, $\Pp\Pp \to \tZq$ (charge conjugation in the final state is implied throughout this Letter). This process is sensitive to a multitude of SM interactions described via the $\PW\PW\cPZ$ triple-gauge coupling, the $\cPqt\cPqt\cPZ$ and $\cPqt\cPqb\PW$ couplings, and the $\cPqb\PW \to \cPqt\cPZ$ scattering amplitude~\cite{Campbell:2013yla}. Because of unitary cancellations in SM \tZq production, the \tZq process might be affected by modified interactions even when neither top quark pair production in association with the $\PZ$ boson (\ttZ), nor inclusive single top quark production would be affected in a visible manner~\cite{Degrande:2018fog}. In addition, modified \tZq production could indicate the presence of flavor changing neutral currents~\cite{AguilarSaavedra:2004wm,Agram:2013koa,Glashow:1970gm}. These unique features, and the addition of complementary information to the global constraints on modified top quark interactions, make the \tZq production cross section an important quantity to measure.

This Letter presents the observation of \tZq production and its cross section measurement, using the leptonic \tZq decay channel in events with three charged leptons, either electrons or muons (including a small contribution from sequential $\PGt$ lepton decays), and at least two additional jets, one of which is identified as originating from a \PQb quark. The analysis is performed using \Pp\Pp\ collision data at $\sqrt{s} = 13\TeV$ collected with the CMS detector in 2016 and 2017, corresponding to an integrated luminosity of 77.4\fbinv. Previous searches for \tZq production by the ATLAS~\cite{ATLAS:tZq2016} and CMS~\cite{CMS:tZq2016} Collaborations at 13\TeV, based on an integrated luminosity of approximately 36\fbinv, resulted in observed significances of 4.2 and 3.7 standard deviations, respectively, from the background-only hypothesis. More than doubling the integrated luminosity by adding the 2017 data, and improvements to the lepton identification techniques and the analysis strategy, significantly increase the sensitivity of the present analysis in comparison to previous searches.

The central feature of the CMS apparatus~\cite{Chatrchyan:2008zzk} is a superconducting solenoid of 6\unit{m} internal diameter, providing a magnetic field of 3.8\unit{T}. Silicon pixel and strip trackers, a lead tungstate crystal electromagnetic calorimeter (ECAL), and a brass and scintillator hadron calorimeter, each composed of a barrel and two endcap sections, reside within the solenoid. Forward calorimeters extend the pseudorapidity ($\eta$) coverage provided by the barrel and endcap detectors. Muons are detected in gas-ionization detectors embedded in the steel flux-return yoke outside the solenoid. Events of interest are recorded with several trigger algorithms~\cite{Khachatryan:2016bia}, requiring the presence of one, two, or three electrons or muons, resulting in an efficiency of almost 100\% for events passing the analysis selection.

Samples of Monte Carlo (MC) simulated events are used to determine the \tZq signal acceptance and to estimate the yields for most of the background processes. Separate MC samples, matching the data taking conditions in 2016 and in 2017, are used.  The \tZq events are simulated with the \MGvATNLO program~\cite{MADGRAPH5, Frixione:2002ik} at next-to-leading order (NLO) in perturbative quantum chromodynamics (QCD). The \MGvATNLO generator is also used for the simulation of the main background processes with at least one top quark ($\cPqt\PH \PW$, $\cPqt\PH \Pq$, $\cPqt \PW \PZ$, \ttV, \ttVV) or three gauge bosons ($\cmsSymbolFace{VVV}$), where $\cmsSymbolFace{V} = \PW$ or $\PZ$, and \PH is the Higgs boson, either at leading order (LO) or at NLO in QCD. The most important of these backgrounds, namely \ttW and \ttZ, are simulated at NLO in QCD. Version 2.2.2 (2.4.2) of \MGvATNLO is used for the simulation of 2016 (2017) collisions. Samples of diboson as well as $\ttbar\PH$ events are produced at NLO precision, using the \POWHEG~v2~\cite{Melia:2011tj,Nason:2013ydw, Nason:2004rx, Frixione:2007vw, Alioli:2010xd} generator.

The NNPDF3.0~\cite{Ball:2014uwa} (NNPDF3.1~\cite{Ball:2017nwa}) parton distribution function (PDF) sets~\cite{Buckley:2014ana} are used for simulation of 2016 (2017) data, with the perturbative order in QCD matching that used in the sample generation. The simulation of parton showering, hadronization, and the underlying event is performed with \PYTHIA 8.212 (8.230)~\cite{Sjostrand:2014zea} for simulated samples matching 2016 (2017) conditions, using the CUETP8M1~\cite{Skands:2014pea,CMS-PAS-GEN-14-001} (CP5~\cite{CP5PAS}) underlying event tune. Double counting of partons generated with \MGvATNLO and \PYTHIA is eliminated using the \textsc{FxFx}~\cite{Frederix:2012ps} (MLM~\cite{Alwall:2007fs}) matching scheme for the NLO (LO) samples.

The effects of additional $\Pp\Pp$ collisions in the same or adjacent bunch crossings (pileup) are taken into account by overlaying each simulated event with a number of inelastic collisions, simulated with \PYTHIA. The generated distribution of the number of events per bunch crossing is matched to that observed in data. Simulated events include a full \GEANTfour-based~\cite{Geant} simulation of the CMS detector and are reconstructed using the same software employed for data.

The particle-flow (PF) algorithm~\cite{Sirunyan:2017ulk} aims to reconstruct and identify each individual particle in an event, with an optimized combination of information from the various elements of the CMS detector, and determine the $\Pp\Pp$ interaction primary vertex (PV)~\cite{CMS:tZq2016}. Reconstructed particles (PF candidates) are classified as charged or neutral hadrons, photons, electrons, or muons.

The PF candidates are clustered into jets using the anti-$\kt$ clustering algorithm~\cite{Cacciari:2008gp} with a distance parameter of 0.4, implemented in the \FASTJET package~\cite{Cacciari:fastjet1,Cacciari:fastjet2}. Jets are required to pass several quality criteria, designed to remove jet candidates that are likely to originate from anomalous energy deposits in the calorimeters~\cite{CMS-PAS-JME-10-003}. Jet energies are corrected for nonlinearity and nonuniformity of the detector response using a combination of simulated samples and $\Pp\Pp$ collision data~\cite{Chatrchyan:2011ds,Khachatryan:2016kdb}. Jets are retained for further analysis if they have a transverse momentum $\pt > 25\GeV$, $\abseta < 5$, and are separated by $\Delta R = \sqrt{\smash[b]{(\Delta\eta)^2+(\Delta\phi)^2}} > 0.4$ from any identified leptons, where $\Delta\eta$ and $\Delta\phi$ are the pseudorapidity and azimuthal angle differences, respectively, between the directions of the jet and the lepton. High-\abseta jets are included to account for forward jets produced by the quark in \tZq events. Because of an increased level of noise in the very forward ECAL region in 2017 data, the minimum \pt threshold for jets in the $2.7 < \abseta < 3.0$ range was raised to 60\GeV for this data set and for the corresponding simulated event samples.

Jets with $\abseta < 2.4$ originating from the hadronization of \PQb quarks are identified with the \textsc{DeepCSV} algorithm~\cite{Sirunyan:2017ezt}. They are considered \PQb tagged, and referred to as \PQb jets, if they pass a working point of this algorithm, which has a typical efficiency of 68\% for correctly identifying \PQb quark jets, with a misidentification probability of 12 (1)\% for \PQc quark (light-flavor) jets.

The missing transverse momentum vector $\ptvecmiss$ is defined as the negative vector \pt sum of all PF candidates in the event, taking into account the jet energy corrections~\cite{CMS-PAS-JME-16-003}. Its magnitude is referred to as \ptmiss.

Electron reconstruction is based on the combination of tracker and ECAL measurements~\cite{Khachatryan:2015hwa}. Each electron candidate must fulfill several quality requirements on the ECAL shower shape and have no more than one missing hit in the tracker. Muons are reconstructed by combining information from the tracker, the muon spectrometers, and the calorimeters in a global fit~\cite{Sirunyan:2018fpa}. Muon candidates must meet criteria on the geometric matching between the signals in different subdetectors and the quality of the global fit. To be considered in the analysis, electron and muon candidates must be consistent with coming from the PV and pass prerequisite selection criteria on their relative isolation, defined as the scalar \pt sum of all PF candidates inside a cone around the lepton, divided by the lepton \pt. The angular radius of the cone in ($\eta$, $\phi$) space is given by $\Delta R(\pt(\ell)) = 10 \GeV/\text{min} \left[ \text{max} \left( \pt(\ell), 50 \GeV  \right) , 200 \GeV \right]$, thus taking into account the increased particle collimation at high \pt values~\cite{Khachatryan:2016kod}. The relative isolation for electrons and muons is required to be below 0.4.

The search crucially depends on efficiently distinguishing leptons originating from the decay of EW bosons from both genuine leptons produced in hadron decays, and photon conversions or jet constituents incorrectly reconstructed as leptons. The first category is referred to as prompt leptons, while the last two are collectively labeled as nonprompt leptons. The reach of the previous analysis of \tZq production by CMS~\cite{CMS:tZq2016} was largely limited by the relative contribution from the nonprompt-lepton background and by the uncertainty in its prediction. Taking this into consideration, gradient boosted decision trees (BDTs) are set up to maximally discriminate between prompt and nonprompt leptons. The BDTs exploit the properties of the jet closest to the lepton in terms of $\Delta R$, the relative isolation defined above, the relative isolation inside a fixed cone size of $\Delta R = 0.3$, the impact parameters of the leptons with respect to the PV, and the lepton \pt and \abseta. Additionally, the BDTs have access to variables related to the ECAL shower shape of electrons and the geometric matching between the silicon tracker and muon system measurements of muons. The BDT discriminants are trained using the \textsc{tmva} package~\cite{Hocker:2007ht}. As a cross-check, fully connected feed-forward neural networks are trained using \textsc{Keras}~\cite{chollet2015keras} with \textsc{Tensorflow}~\cite{tensorflow2015-whitepaper} as backend, which lead to nearly identical performance.

A stringent requirement is placed on the BDT output, resulting in a selection efficiency of 85~(92)\% per prompt electron (muon) with $\pt > 25$\GeV passing the prerequisite selection criteria, as measured in simulated \tZq events. The corresponding misidentification probability for simulated nonprompt leptons from \ttbar events is about 1.5\%. Compared to the non-BDT-based lepton identification used in the previous analysis~\cite{CMS:tZq2016}, the selection efficiency for prompt electrons (muons) improves by up to 12 (8)\%, while rejecting more nonprompt leptons by a factor of approximately 2 (8) in simulated events.

The analysis uses two definitions for the lepton selection. Leptons that pass the aforementioned BDT selection criteria are referred to as ``tight leptons". ``Loose leptons'' are the combined set of tight leptons and leptons that pass, on top of the prerequisite ones, loose selection criteria based on the attributes of the closest jet, and in the case of electrons, on a multivariate discriminant based on the ECAL shower shape~\cite{Khachatryan:2015hwa}. The loose selection is optimized to provide a reliable prediction of the  nonprompt-lepton background, as explained below.

To be considered in the analysis, events must contain exactly three loose leptons, two of which form a pair of opposite sign and same flavor (OSSF) with an invariant mass within a window of $30\GeV$ width centered on the world-average $\PZ$ boson mass~\cite{PDG}. All three selected leptons must pass the tight selection requirements in order for the event to enter the final selection. Events in which at least one of the leptons fails to pass the tight criteria are used to estimate the nonprompt-lepton background. The three leptons, ordered from highest to lowest \pt, are required to have \pt values greater than 25, 15, and 10\GeV, respectively.

Events are divided into three categories, collectively referred to as signal regions (SRs), based on the number of jets they contain. Events with a total of two or three jets, exactly one of which is \PQb tagged, make up SR-2/3j-1b, which contains most \tZq events. Events with four or more jets, exactly one of which is \PQb tagged, form SR-4j-1b, while SR-2b contains events with two or more \PQb-tagged jets. Events without \PQb-tagged jets, or with one \PQb-tagged jet and no additional jets, have a very low signal-to-background ratio and are rejected.

In each of these categories, a dedicated BDT is trained to extract the \tZq signal from the total background on several discriminating variables, using the \textsc{tmva} package. Half of the simulated signal and background events are randomly selected and used for training, while the rest are used for testing. The most significant difference between  the \tZq signal and background events is the tendency of the \tZq events to have a forward jet. Simulated signal events show that at least one jet has a high \abseta value and produces a large dijet invariant mass when combined with another jet in the event. The \PQb-tagged jet yielding the invariant mass closest to the top quark mass~\cite{PDG}, when combined with the \ptvecmiss and the lepton ($\ell_{\PW}$) not forming the \PZ boson candidate, is considered as originating from the top quark decay. The remaining jet with the highest \pt in the event, typically found in the forward region of the detector, is labeled the ``recoiling jet".

The following variables are used to construct the BDT discriminants: the \abseta of the recoiling jet, the maximum dijet invariant mass among all pairs of jets in the event, the sums of leptonic and hadronic transverse momenta in the event, the transverse mass of the combination of $\vec{\pt}^{\ell_{\PW}}$ and \ptvecmiss ($\sqrt{\smash[b]{2\pt^{\ell_{\PW}}\ptmiss\{1-\cos[\Delta\phi(\vec{\pt}^{\ell_{\PW}}, \ptvecmiss)]\}}}$), the \abseta of $\ell_{\PW}$ multiplied by its charge, the highest \textsc{DeepCSV} discriminant value among all jets in the event, the maximum azimuthal separation between any two of the leptons, and the minimum $\Delta R$ separation between any lepton and \PQb-tagged jet. For events in SR-2/3j-1b, the maximum \pt of any dijet system is used as an additional input variable, while for SR-4j-1b and SR-2b, the invariant mass of the three-lepton system and the \abseta of the most forward jet are included to improve the BDT performance. In addition, for SR-4j-1b, the $\Delta R$ separation between $\ell_{\PW}$ and the \PQb-tagged jet, and between this jet and the recoiling jet are added as BDT inputs. The modeling of each BDT input variable in simulation was validated in data. The \tZq cross section measurement and signal significance are obtained from a binned maximum-likelihood fit to the distributions of the resulting BDT discriminants.

The background contributions to the three SRs are divided into two groups: those that have three or more prompt leptons, and those containing at least one nonprompt lepton. The contribution from the former group is estimated from simulation, while the contribution from the latter is predicted directly from data.

The largest background in SR-2/3j-1b comes from \WZ production. It is estimated from simulation and its normalization is measured in a control data sample enriched in \WZ events. The control sample consists of events passing the same selection as the SR events, but with no requirements on the number of jets and with an explicit veto on events with a \PQb-tagged jet. Additionally, $\ptmiss > 50\GeV$ is required. A prior uncertainty of 10\% is assumed in the \WZ normalization, and an additional extrapolation uncertainty of 8\% is assigned to \WZ events with one or more \PQb jets. The latter uncertainty is based on dedicated studies in data events enriched in Z bosons accompanied by the gluon splitting process yielding a pair of \PQb jets.

The \ttZ process has a large branching fraction to three prompt leptons and is the dominant background in SR-4j-1b and SR-2b. The contribution from \ttZ events is estimated from simulation, and its shape and normalization are further constrained in the final fit via the bins at low BDT values, whose contents are dominated by \ttZ events. A prior uncertainty of 15\% is assigned to the \ttZ normalization.

Other processes involving a top quark pair or a single top quark produced in association with additional particles ($\ttbar\text{X}\,/\,\cPqt\text{X}$) also contribute to the background. These contributions are estimated using simulation and mainly come from \tth , \ttW, and \tWZ production. These processes are normalized to their predicted cross sections, accounting for theoretical uncertainties.

Events with four or more prompt leptons enter the selection if at least one of the leptons fails to be identified. This background consists mainly of \ZZ and \ttZ events, and is largely reduced by applying a veto on the presence of a loose fourth lepton. The \ZZ background normalization is constrained via a control data sample of four-lepton events, in which there are two OSSF pairs with invariant masses close to that of the \PZ boson. A prior uncertainty of 10\% is assumed in the normalization of \ZZ.

Internal and external conversions of photons could result in additional leptons in an event. This typically occurs through an asymmetric conversion, in which one of the leptons coming from the conversion has very low \pt and fails to be reconstructed. This background (\XgammaStar, where $\text{X}$ stands for any combination of massive EW bosons or top quarks), dominated by \ttgamma and \Zgamma events, is obtained from simulation. A control data sample of three-lepton events is enriched in \Zgamma events by requiring the invariant mass of the three-lepton system to be within a 30\GeV window centered at the nominal \PZ boson mass, while no lepton pair is allowed to have an invariant mass within this window. This control sample is used to validate the simulation of conversions, and the data and simulation were found to agree within the uncertainties.

The final background contribution with three prompt leptons comes from rare processes involving multiple massive EW bosons. Such processes have very small cross sections and branching fractions to multiple leptons, so their contribution is minimal. This background is estimated using simulation scaled to the respective predicted cross sections, taking into account theoretical uncertainties.

Events with nonprompt leptons that enter the SRs mainly consist of \ttbar and Drell--Yan events with an additional nonprompt lepton. Their contribution is estimated directly from data using the ``tight-to-loose" ratio method, as described in Ref.~\cite{Khachatryan:2016kod}. The probability for a loose nonprompt lepton to pass the tight selection requirements is measured as a function of its \pt and \abseta in a control data sample of QCD multijet events, rich in nonprompt leptons. The measured probability is then applied to data events in which one or more leptons fail the tight selection, while passing the loose selection. The method is validated in both simulation and control data samples enriched in \ttbar and Drell--Yan events. The agreement between the predicted and observed yields is found to be within 30\% in the most relevant kinematic distributions, and an uncertainty of 30\% is therefore assigned to the prediction of this background. Owing to the high performance of the BDT-based lepton selection used in the analysis, the contribution of this background is small compared to that with three prompt leptons.

A number of sources of experimental uncertainty affect each of the simulated samples. These sources include pileup modeling, jet energy scale, \PQb tagging, trigger and lepton identification efficiencies, $\ptmiss$ resolution, and the integrated luminosity. Theoretical uncertainties in the fixed-order cross section calculations used to normalize the simulated samples are an additional source of systematic uncertainty. The effects of each of these sources, except the ones associated with the integrated luminosity and trigger efficiency, vary across the BDT distribution.

The uncertainty in the simulated distribution of the number of events per bunch crossing is estimated by varying the total $\Pp\Pp$ inelastic cross section by $\pm$4.6\%~\cite{Sirunyan:2018nqx}. This causes variations in the simulated event yields of 0.7--5.0\% across the BDT bins. The integrated luminosity, used to normalize the simulated event yields, is measured with a precision of 2.5\,(2.3)\% in the data collected in 2016~\cite{CMS-PAS-LUM-17-001} (2017~\cite{CMS-PAS-LUM-17-004}).

The uncertainty from the jet energy scale is estimated by varying the scale up and down within its uncertainty for all jets in the event~\cite{Khachatryan:2016kdb}. The effect of this variation is propagated through all steps of the analysis. The resulting variations across the BDT bins range from 1.5--15\%\,(1.8--38\%) in 2016\,(2017) data. Corrections applied to account for the differences between data and simulation in the \PQb tagging efficiency and misidentification rate lead to an uncertainty of 0.1--4.4\% in the simulated event yields per bin.

The trigger efficiency is measured by selecting events with three leptons in an unbiased data sample, triggered on the $\ptmiss$ or hadronic activity in the event. Statistical uncertainties in this measurement lead to a 2\% uncertainty in the trigger efficiency. The lepton identification efficiencies are measured in data using the ``tag-and-probe" technique~\cite{Sirunyan:2018fpa,Khachatryan:2015hwa}, and corresponding corrections are applied to simulation. For muons the efficiency corrections are typically around 1\%, and go up to 5\% in the forward region. For electrons the typical efficiency corrections are 5\%, and are as high as 20\% for forward, low-\pt electrons. Uncertainties in the efficiency measurements lead to a total uncertainty of 2.5--4.9\% in the simulated event yields per BDT bin.

Uncertainties from the choice of the renormalization and factorization scales used in simulation are assessed by simultaneously varying these scales up and down by a factor of two, resulting in uncertainties of 0.8--9.6\% in the simulated yields per BDT bin. The limited knowledge of the proton PDFs is taken into account using  a set of NNPDF3.0 (NNPDF3.1) replicas~\cite{Butterworth:2015oua} in the simulation of 2016 (2017) collisions and leads to uncertainties of 0.04--1.4\%. These theoretical uncertainties are taken into account for all simulated samples, and cause changes in both the predicted cross section and the detector acceptance for simulated events, which are treated independently. For \WZ, \ttZ, \ZZ, and \tZq production, theoretical uncertainties in the cross section are not taken into consideration, and prior nuisance parameters are assigned to their normalizations that are constrained by data. For all other processes, such as \ttW, \tth, $\cPqt \PW \PZ$, and triple gauge boson production, theoretical uncertainties in the predicted cross sections are included. Similarly, the uncertainty in the parton shower simulation is estimated by varying the renormalization scales for both initial- and final-state radiation up and down by a factor of 2~\cite{Skands:2014pea}. This source of uncertainty is only considered for simulated \tZq and \ttZ processes and ranges from 0.1--6.5\% (0.3--7.3\%) across the BDT bins for the description of initial-(final-)state radiation.

\ifthenelse{\boolean{cms@external}}
{\begin{figure*}[thbp!]
\centering
\includegraphics[width=.325\textwidth]{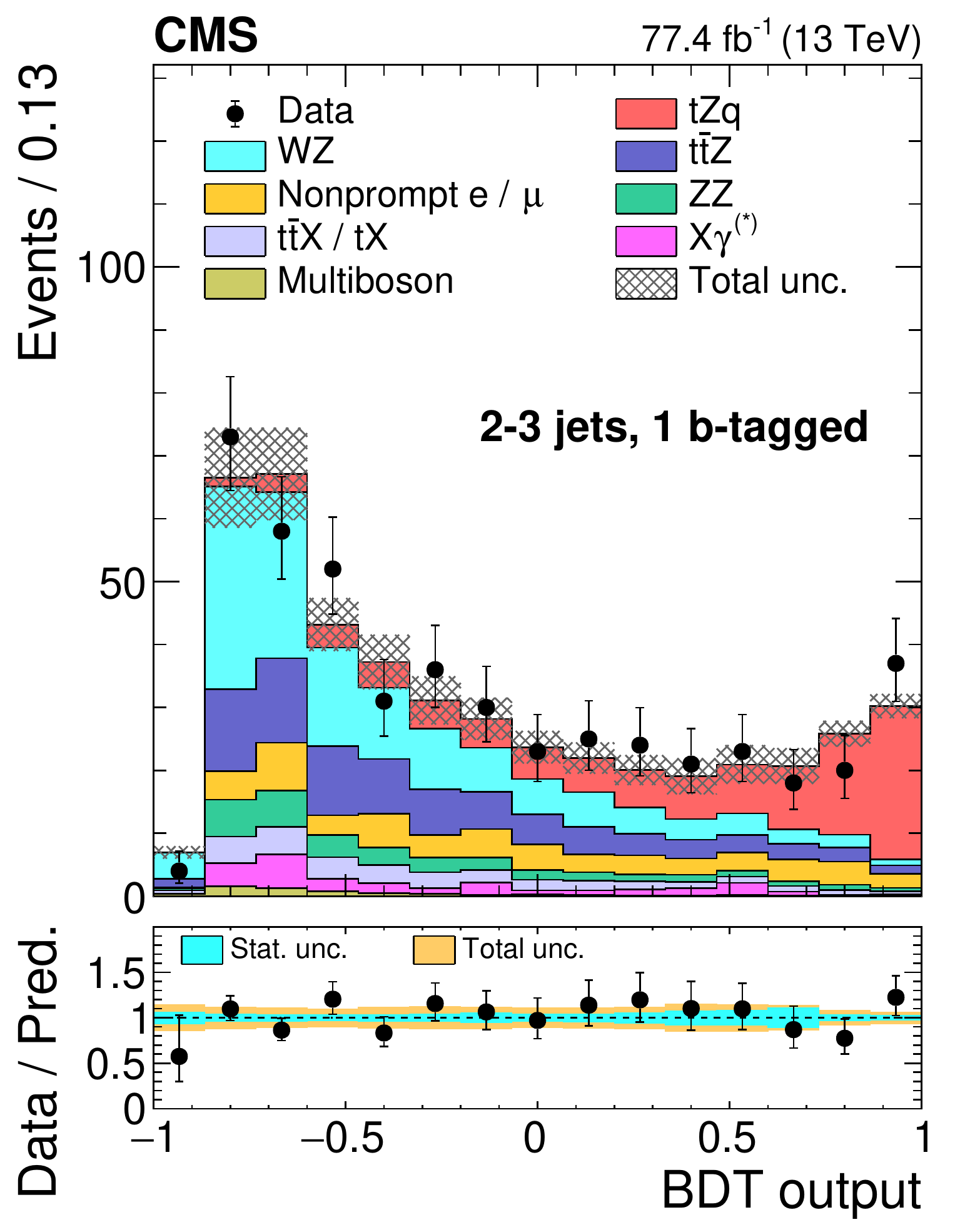}
\includegraphics[width=.325\textwidth]{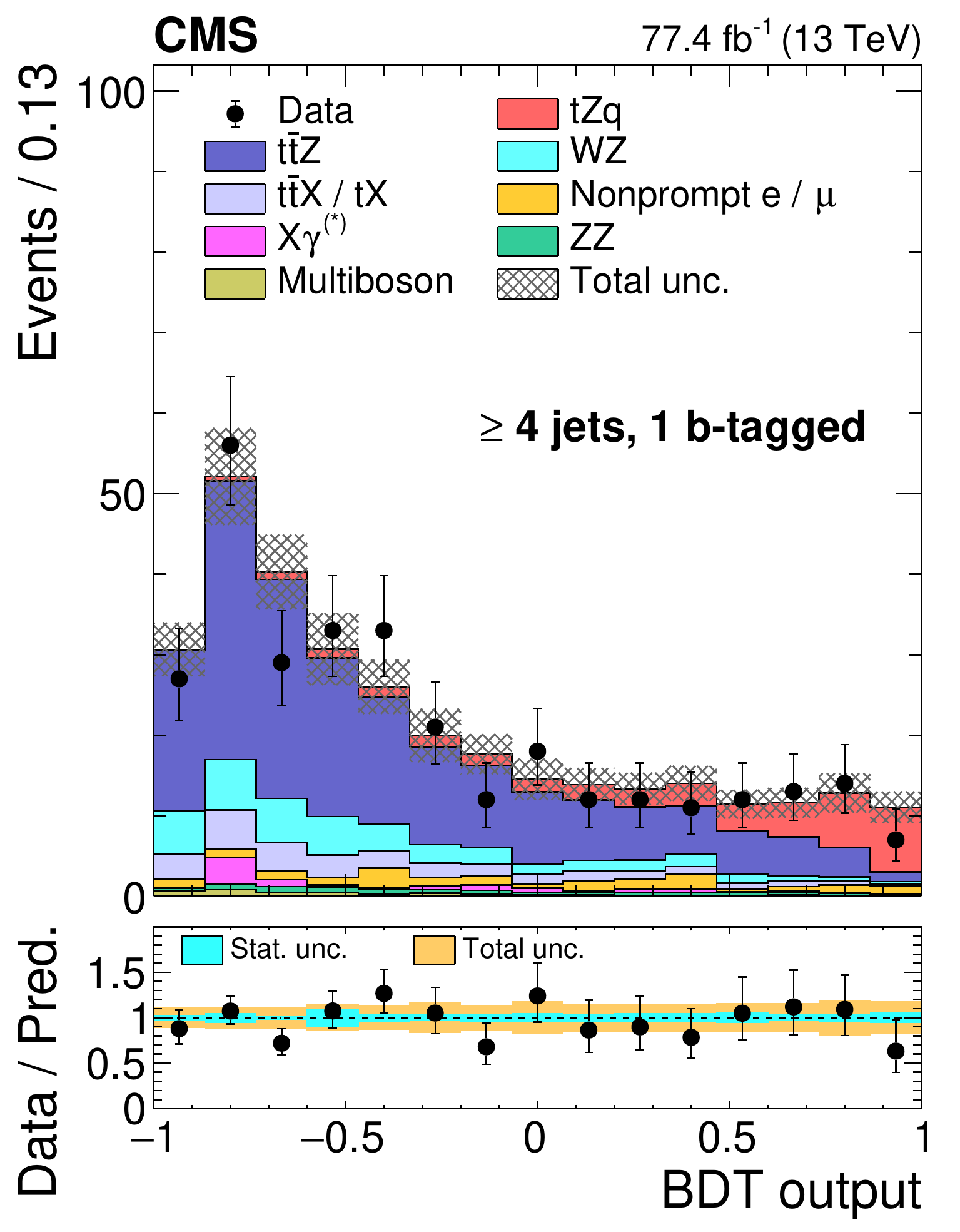}
\includegraphics[width=.325\textwidth]{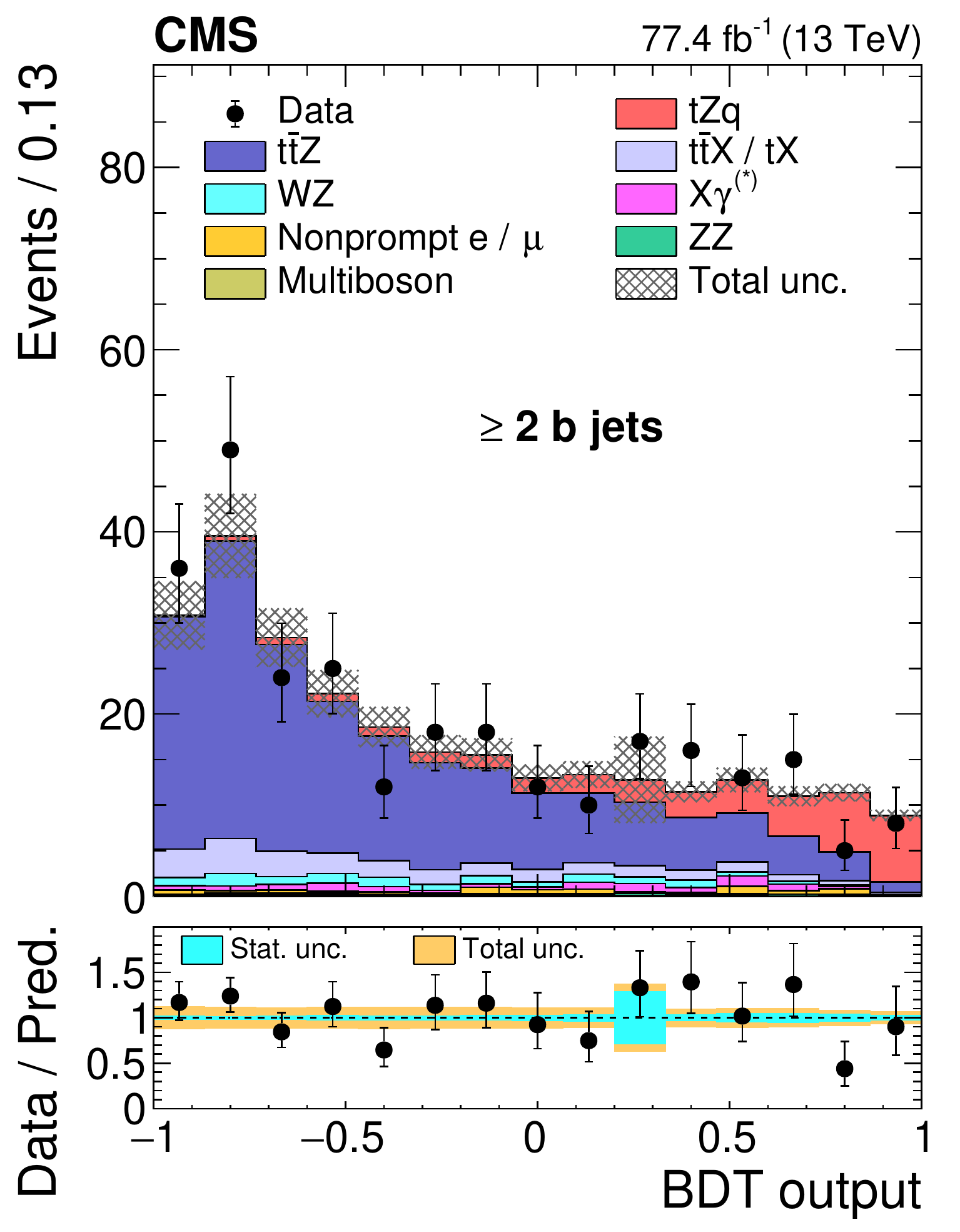}
\caption{Observed (points) and post-fit expected (shaded histograms) BDT distributions for events in SR-2/3j-1b (left), SR-4j-1b (middle), and SR-2b (right). The vertical bars on the points represent the statistical uncertainties in data. The hatched regions show the total uncertainties in the background. The lower panels display the ratio of the observed data to the predictions, including the \tZq signal, with inner and outer shaded bands, respectively, representing the statistical and total uncertainties in the predictions.}
\label{fig:observedYields}
\end{figure*}
}{}

A simultaneous binned maximum-likelihood fit to the BDT distributions, and to the event yields in the \WZ and \ZZ control regions, is performed to measure the \tZq signal strength. The best fit value of the signal strength and the 68\% confidence interval are extracted following the procedure described in Section 3.2 of Ref.~\cite{Khachatryan:2014jba}. All sources of systematic uncertainties are taken into account as nuisance parameters in the fit. The appropriate correlation pattern of the nuisance parameters between the 2016 and 2017 data sets is taken into account; the nuisance parameters associated with the integrated luminosity, \PQb tagging, trigger efficiency, and jet energy scale modeling are considered to be fully uncorrelated between the two data taking periods, while all others are considered to be fully correlated.

The observed (expected) statistical significance of the signal is determined using the asymptotic approximation of the distribution of the profile likelihood test statistic~\cite{Cowan:2010js,ATLAS:1379837} and found to be 8.2 (7.7) standard deviations from the background-only hypothesis. The analyses based on the 2016 and 2017 data sets result in observed (expected) signal significances of 7.2 (5.7) and 5.4 (6.0) standard deviations, respectively. The \tZq cross section is measured to be
\begin{linenomath}
\begin{equation}
\sigma(\Pp\Pp \to \tZq \to \cPqt \ell^{+} \ell^{-} \cPq) = 111 \pm 13 \stat \, _{-9}^{+11} \syst \unit{fb},
\end{equation}
\end{linenomath}
where $\ell$ refers to an electron, muon, or $\PGt$ lepton, for invariant masses of the dilepton pair larger than 30\GeV. The theoretical cross section in the same fiducial volume is $\sigma^\text{SM}(\Pp\Pp \to \tZq \to \cPqt \ell^{+} \ell^{-} \cPq) = 94.2 \pm 3.1\unit{fb}$, which is computed at NLO in perturbative QCD using the NNPDF3.0 PDF set in the five-flavor scheme~\cite{CMS:tZq2016}. The measured signal strength is
\begin{linenomath}
\ifthenelse{\boolean{cms@external}}{
\begin{multline}
\mu = \frac{\sigma(\Pp\Pp \to\tZq \to \cPqt \ell^{+} \ell^{-} \cPq)}{\sigma^\text{SM}(\Pp\Pp \to\tZq \to \cPqt \ell^{+} \ell^{-} \cPq)} \\
= 1.18 \, _{-0.13}^{+0.14} \stat \, _{-0.10}^{+0.11} \syst \, _{-0.04}^{+0.04} \thy,
\end{multline}}
{
\begin{equation}
\mu = \frac{\sigma(\Pp\Pp \to\tZq \to \cPqt \ell^{+} \ell^{-} \cPq)}{\sigma^\text{SM}(\Pp\Pp \to\tZq \to \cPqt \ell^{+} \ell^{-} \cPq)} = 1.18 \, _{-0.13}^{+0.14} \stat \, _{-0.10}^{+0.11} \syst \, _{-0.04}^{+0.04} \thy,
\end{equation}
}
\end{linenomath}
consistent with the SM expectation. The quoted theoretical uncertainty stems from the uncertainty in $\sigma^\text{SM}(\Pp\Pp \to\tZq \to \cPqt \ell^{+} \ell^{-} \cPq)$. The signal strengths measured separately in the 2016 and 2017 data sets are found to be consistent with the combined measurement, and are $1.36 \allowbreak \, _{-0.20}^{+0.22} \stat \allowbreak \,_{-0.12}^{+0.14} \syst \, _{-0.04}^{+0.04} \thy$ and $1.03 \, _{-0.17}^{+0.18} \stat \, _{-0.12}^{+0.14} \syst \, _{-0.03}^{+0.03} \thy$, respectively. The systematic uncertainties with the largest contribution to the final measurement are those associated with the nonprompt-lepton background prediction, the lepton selection efficiency, the modeling of final-state radiation, and the jet energy scale. The uncertainty in the jet energy scale is constrained by the fit to be approximately twice smaller than its input value, while the other aforementioned uncertainties are not significantly constrained. A table showing the impact of the most important uncertainty sources on the measurement is presented in \suppMaterial.

The observed and expected BDT distributions in each of the SRs are shown in Fig.~\ref{fig:observedYields}. A table with the observed and expected event yields in the SRs and the control regions, and the distributions in SR-2/3j-1b of the maximum dijet mass among all pairs of jets in the event, the \abseta of the recoiling jet, and the reconstructed \PZ boson \pt in events with a BDT discriminant value greater than 0.5 can be found in \suppMaterial. The first two observables are the most discriminant input variables to the BDTs used for signal extraction, while the last one is highly sensitive to the presence of new physics phenomena. The distribution of the number of jets in the event in the \WZ and \ZZ control regions can also be found in \suppMaterial.

\ifthenelse{\boolean{cms@external}}{}
{\begin{figure*}[htb!]
\centering
\includegraphics[width=.325\textwidth]{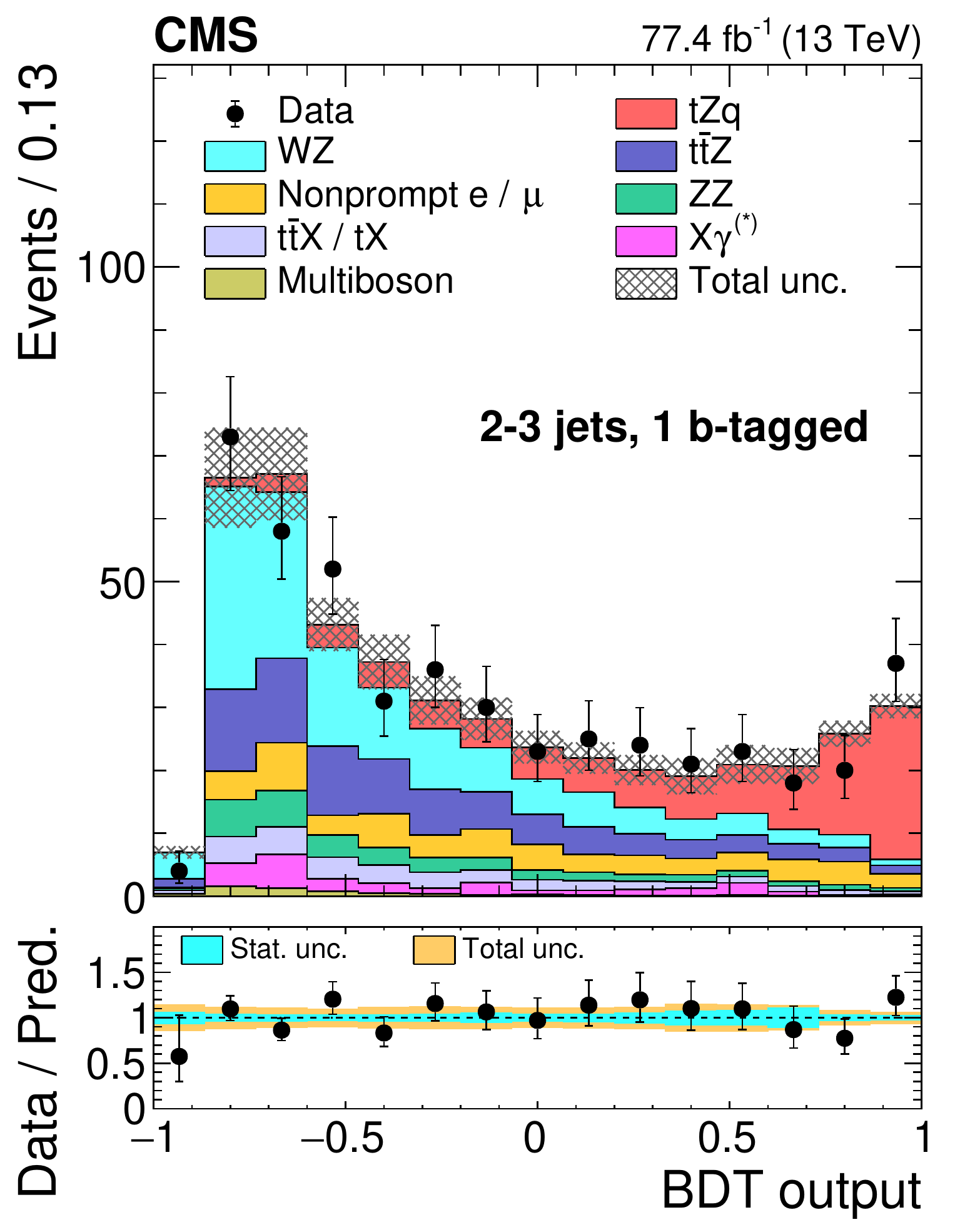}
\includegraphics[width=.325\textwidth]{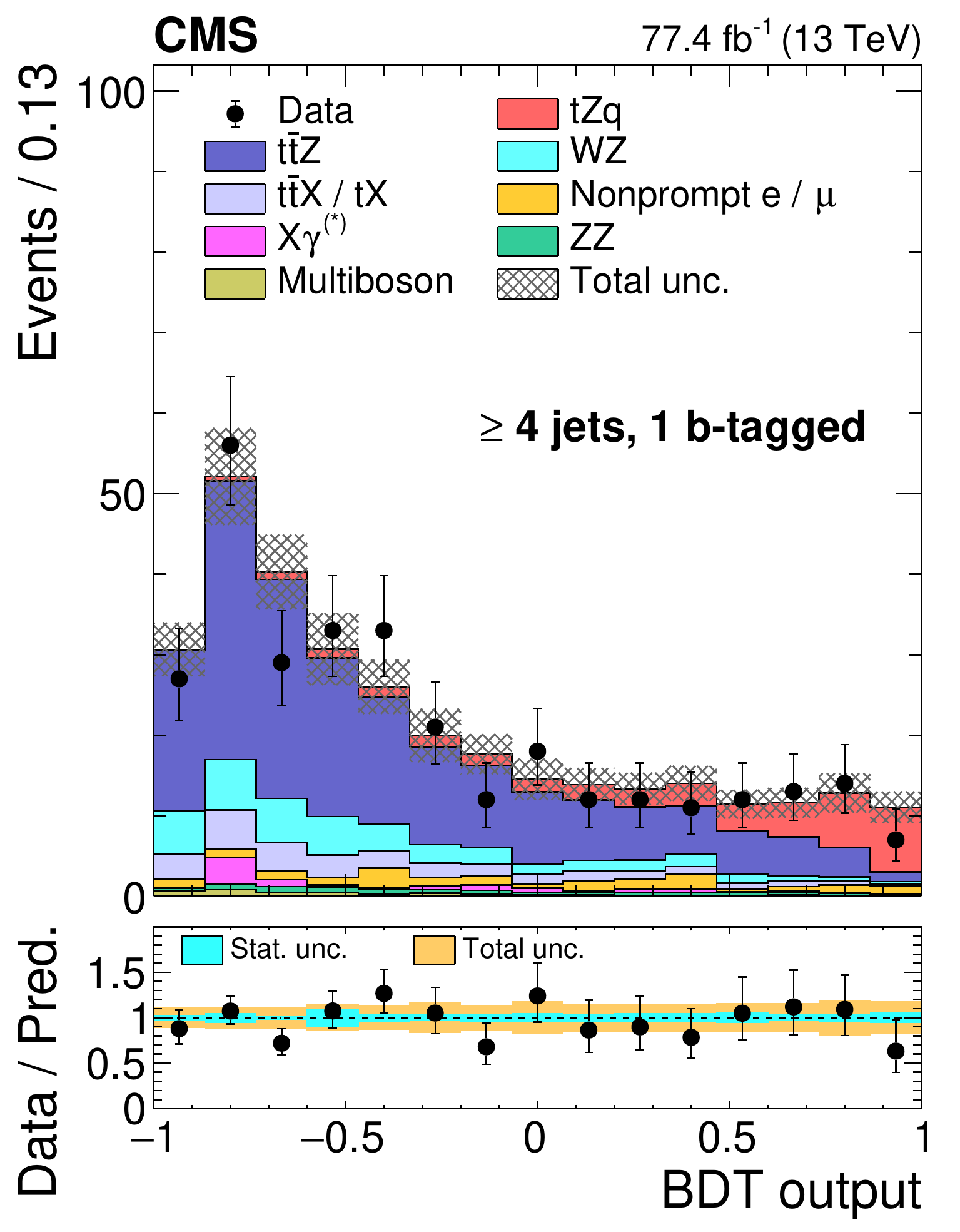}
\includegraphics[width=.325\textwidth]{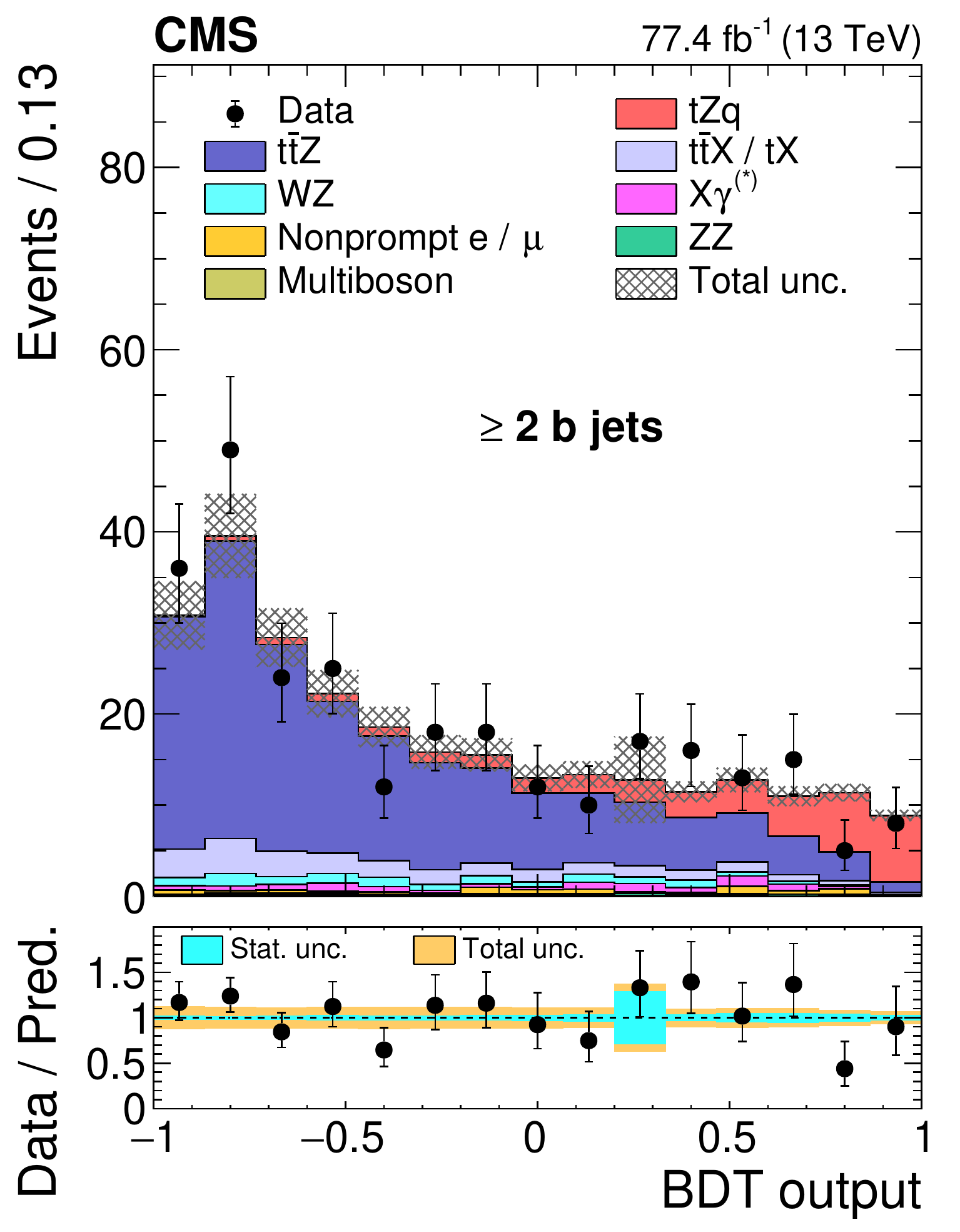}
\caption{Observed (points) and post-fit expected (shaded histograms) BDT distributions for events in SR-2/3j-1b (left), SR-4j-1b (middle), and SR-2b (right). The vertical bars on the points represent the statistical uncertainties in data. The hatched regions show the total uncertainties in the background. The lower panels display the ratio of the observed data to the predictions, including the \tZq signal, with inner and outer shaded bands, respectively, representing the statistical and total uncertainties in the predictions.}
\label{fig:observedYields}
\end{figure*}}

In summary, we have reported the observation of single top quark production in association with a $\PZ$ boson and a quark, \tZq, using the leptonic \tZq decay mode. The \tZq signal is observed with a significance of well over five standard deviations. The \tZq production cross section is measured to be $\sigma(\Pp\Pp \to \tZq \to \cPqt \ell^{+} \ell^{-} \cPq) = 111 \pm 13 \stat \, _{-9}^{+11} \syst \unit{fb}$, where $\ell$ refers to an electron, muon, or $\PGt$ lepton, for dilepton invariant masses in excess of 30\GeV, in agreement with the standard model prediction.

\begin{acknowledgments}
We congratulate our colleagues in the CERN accelerator departments for the excellent performance of the LHC and thank the technical and administrative staffs at CERN and at other CMS institutes for their contributions to the success of the CMS effort. In addition, we gratefully acknowledge the computing centers and personnel of the Worldwide LHC Computing Grid for delivering so effectively the computing infrastructure essential to our analyses. Finally, we acknowledge the enduring support for the construction and operation of the LHC and the CMS detector provided by the following funding agencies: BMBWF and FWF (Austria); FNRS and FWO (Belgium); CNPq, CAPES, FAPERJ, FAPERGS, and FAPESP (Brazil); MES (Bulgaria); CERN; CAS, MoST, and NSFC (China); COLCIENCIAS (Colombia); MSES and CSF (Croatia); RPF (Cyprus); SENESCYT (Ecuador); MoER, ERC IUT, and ERDF (Estonia); Academy of Finland, MEC, and HIP (Finland); CEA and CNRS/IN2P3 (France); BMBF, DFG, and HGF (Germany); GSRT (Greece); NKFIA (Hungary); DAE and DST (India); IPM (Iran); SFI (Ireland); INFN (Italy); MSIP and NRF (Republic of Korea); MES (Latvia); LAS (Lithuania); MOE and UM (Malaysia); BUAP, CINVESTAV, CONACYT, LNS, SEP, and UASLP-FAI (Mexico); MOS (Montenegro); MBIE (New Zealand); PAEC (Pakistan); MSHE and NSC (Poland); FCT (Portugal); JINR (Dubna); MON, RosAtom, RAS, RFBR, and NRC KI (Russia); MESTD (Serbia); SEIDI, CPAN, PCTI, and FEDER (Spain); MOSTR (Sri Lanka); Swiss Funding Agencies (Switzerland); MST (Taipei); ThEPCenter, IPST, STAR, and NSTDA (Thailand); TUBITAK and TAEK (Turkey); NASU and SFFR (Ukraine); STFC (United Kingdom); DOE and NSF (USA).
\end{acknowledgments}

\bibliography{auto_generated}

\providecommand{\href}[2]{#2}\begingroup\raggedright\begin{thebibliography}{10}%
\makeatletter
\providecommand{\hrefCMSnoop }[0]{\@secondoftwo}%
\makeatother
\providecommand{\doi}{\texttt{doi:}\begingroup \urlstyle{tt}\Url}

\bibitem{Campbell:2013yla}
\hrefCMSnoop {}{J.~Campbell, R.~K. Ellis, and R.~R{\"o}ntsch, ``{Single top
  production in association with a Z boson at the LHC}'',} \textit{ Phys. Rev.
  D} \textbf{ 87} (2013) 114006,
  \href{http://dx.doi.org/10.1103/PhysRevD.87.114006}{\doi{10.1103/PhysRevD.87.114006}},
\href{http://www.arXiv.org/abs/1302.3856}{\texttt{arXiv:1302.3856}}.

\bibitem{Degrande:2018fog}
C.~Degrande\hrefCMSnoop {}{ {et~al.}, ``{Single-top associated production with
  a Z or H boson at the LHC: the SMEFT interpretation}'',} \textit{ JHEP}
  \textbf{ 10} (2018) 005,
  \href{http://dx.doi.org/10.1007/JHEP10(2018)005}{\doi{10.1007/JHEP10(2018)005}},
\href{http://www.arXiv.org/abs/1804.07773}{\texttt{arXiv:1804.07773}}.

\bibitem{AguilarSaavedra:2004wm}
\href
  {http://www.actaphys.uj.edu.pl/fulltext?series=Reg&vol=35&page=2695}{J.~A.
  Aguilar-Saavedra, ``{Top flavor-changing neutral interactions: theoretical
  expectations and experimental detection}'',} \textit{ Acta Phys. Polon. B}
  \textbf{ 35} (2004) 2695,
\href{http://www.arXiv.org/abs/hep-ph/0409342}{\texttt{arXiv:hep-ph/0409342}}.

\bibitem{Agram:2013koa}
J.-L. Agram\hrefCMSnoop {}{ {et~al.}, ``{Probing top anomalous couplings at the
  LHC with trilepton signatures in the single top mode}'',} \textit{ Phys.
  Lett. B} \textbf{ 725} (2013) 123,
  \href{http://dx.doi.org/10.1016/j.physletb.2013.06.052}{\doi{10.1016/j.physletb.2013.06.052}},
\href{http://www.arXiv.org/abs/1304.5551}{\texttt{arXiv:1304.5551}}.

\bibitem{Glashow:1970gm}
\hrefCMSnoop {}{S.~L. Glashow, J.~Iliopoulos, and L.~Maiani, ``{Weak
  interactions with lepton-hadron symmetry}'',} \textit{ Phys. Rev. D} \textbf{
  2} (1970) 1285,
\href{http://dx.doi.org/10.1103/PhysRevD.2.1285}{\doi{10.1103/PhysRevD.2.1285}}.

\bibitem{ATLAS:tZq2016}
\hrefCMSnoop {}{{ATLAS Collaboration}, ``{Measurement of the production
  cross-section of a single top quark in association with a Z boson in
  proton-proton collisions at 13 TeV with the ATLAS detector}'',} \textit{
  Phys. Lett. B} \textbf{ 780} (2018) 557,
  \href{http://dx.doi.org/10.1016/j.physletb.2018.03.023}{\doi{10.1016/j.physletb.2018.03.023}},
\href{http://www.arXiv.org/abs/1710.03659}{\texttt{arXiv:1710.03659}}.

\bibitem{CMS:tZq2016}
\hrefCMSnoop {}{{CMS Collaboration}, ``{Measurement of the associated
  production of a single top quark and a Z boson in pp collisions at $\sqrt{s}
  =$ 13 TeV}'',} \textit{ Phys. Lett. B} \textbf{ 779} (2018) 358,
  \href{http://dx.doi.org/10.1016/j.physletb.2018.02.025}{\doi{10.1016/j.physletb.2018.02.025}},
\href{http://www.arXiv.org/abs/1712.02825}{\texttt{arXiv:1712.02825}}.

\bibitem{Chatrchyan:2008zzk}
\hrefCMSnoop {}{{CMS Collaboration}, ``The {CMS} experiment at the {CERN}
  {LHC}'',} \textit{ JINST} \textbf{ 3} (2008) S08004,
\href{http://dx.doi.org/10.1088/1748-0221/3/08/S08004}{\doi{10.1088/1748-0221/3/08/S08004}}.

\bibitem{Khachatryan:2016bia}
\hrefCMSnoop {}{{CMS Collaboration}, ``{The CMS trigger system}'',} \textit{
  JINST} \textbf{ 12} (2017) P01020,
  \href{http://dx.doi.org/10.1088/1748-0221/12/01/P01020}{\doi{10.1088/1748-0221/12/01/P01020}},
\href{http://www.arXiv.org/abs/1609.02366}{\texttt{arXiv:1609.02366}}.

\bibitem{MADGRAPH5}
J.~Alwall\hrefCMSnoop {}{ {et~al.}, ``{The automated computation of tree-level
  and next-to-leading order differential cross sections, and their matching to
  parton shower simulations}'',} \textit{ JHEP} \textbf{ 07} (2014) 079,
  \href{http://dx.doi.org/10.1007/JHEP07(2014)079}{\doi{10.1007/JHEP07(2014)079}},
\href{http://www.arXiv.org/abs/1405.0301}{\texttt{arXiv:1405.0301}}.

\bibitem{Frixione:2002ik}
\hrefCMSnoop {}{S.~Frixione and B.~R. Webber, ``{Matching NLO QCD computations
  and parton shower simulations}'',} \textit{ JHEP} \textbf{ 06} (2002) 029,
  \href{http://dx.doi.org/10.1088/1126-6708/2002/06/029}{\doi{10.1088/1126-6708/2002/06/029}},
\href{http://www.arXiv.org/abs/hep-ph/0204244}{\texttt{arXiv:hep-ph/0204244}}.

\bibitem{Melia:2011tj}
\hrefCMSnoop {}{T.~Melia, P.~Nason, R.~R{\"o}ntsch, and G.~Zanderighi,
  ``{$\PW^+ \PW^-$, $\PW \PZ$ and $\PZ \PZ$ production in the POWHEG BOX}'',}
  \textit{ JHEP} \textbf{ 11} (2011) 078,
  \href{http://dx.doi.org/10.1007/JHEP11(2011)078}{\doi{10.1007/JHEP11(2011)078}},
\href{http://www.arXiv.org/abs/1107.5051}{\texttt{arXiv:1107.5051}}.

\bibitem{Nason:2013ydw}
\hrefCMSnoop {}{P.~Nason and G.~Zanderighi, ``{$\PW^+ \PW^-$, $\PW \PZ$ and
  $\PZ \PZ$ production in the POWHEG-BOX-V2}'',} \textit{ Eur. Phys. J. C}
  \textbf{ 74} (2014) 2702,
  \href{http://dx.doi.org/10.1140/epjc/s10052-013-2702-5}{\doi{10.1140/epjc/s10052-013-2702-5}},
\href{http://www.arXiv.org/abs/1311.1365}{\texttt{arXiv:1311.1365}}.

\bibitem{Nason:2004rx}
\hrefCMSnoop {}{P.~Nason, ``{A new method for combining NLO QCD with shower
  Monte Carlo algorithms}'',} \textit{ JHEP} \textbf{ 11} (2004) 040,
  \href{http://dx.doi.org/10.1088/1126-6708/2004/11/040}{\doi{10.1088/1126-6708/2004/11/040}},
\href{http://www.arXiv.org/abs/hep-ph/0409146}{\texttt{arXiv:hep-ph/0409146}}.

\bibitem{Frixione:2007vw}
\hrefCMSnoop {}{S.~Frixione, P.~Nason, and C.~Oleari, ``{Matching NLO QCD
  computations with parton shower simulations: the POWHEG method}'',} \textit{
  JHEP} \textbf{ 11} (2007) 070,
  \href{http://dx.doi.org/10.1088/1126-6708/2007/11/070}{\doi{10.1088/1126-6708/2007/11/070}},
\href{http://www.arXiv.org/abs/0709.2092}{\texttt{arXiv:0709.2092}}.

\bibitem{Alioli:2010xd}
\hrefCMSnoop {}{S.~Alioli, P.~Nason, C.~Oleari, and E.~Re, ``{A general
  framework for implementing NLO calculations in shower Monte Carlo programs:
  the POWHEG BOX}'',} \textit{ JHEP} \textbf{ 06} (2010) 043,
  \href{http://dx.doi.org/10.1007/JHEP06(2010)043}{\doi{10.1007/JHEP06(2010)043}},
\href{http://www.arXiv.org/abs/1002.2581}{\texttt{arXiv:1002.2581}}.

\bibitem{Ball:2014uwa}
\hrefCMSnoop {}{{NNPDF} Collaboration, ``{Parton distributions for the LHC Run
  II}'',} \textit{ JHEP} \textbf{ 04} (2015) 040,
  \href{http://dx.doi.org/10.1007/JHEP04(2015)040}{\doi{10.1007/JHEP04(2015)040}},
\href{http://www.arXiv.org/abs/1410.8849}{\texttt{arXiv:1410.8849}}.

\bibitem{Ball:2017nwa}
\hrefCMSnoop {}{{NNPDF} Collaboration, ``{Parton distributions from
  high-precision collider data}'',} \textit{ Eur. Phys. J. C} \textbf{ 77}
  (2017) 663,
  \href{http://dx.doi.org/10.1140/epjc/s10052-017-5199-5}{\doi{10.1140/epjc/s10052-017-5199-5}},
\href{http://www.arXiv.org/abs/1706.00428}{\texttt{arXiv:1706.00428}}.

\bibitem{Buckley:2014ana}
A.~Buckley\hrefCMSnoop {}{ {et~al.}, ``{LHAPDF6: parton density access in the
  LHC precision era}'',} \textit{ Eur. Phys. J. C} \textbf{ 75} (2015) 132,
  \href{http://dx.doi.org/10.1140/epjc/s10052-015-3318-8}{\doi{10.1140/epjc/s10052-015-3318-8}},
\href{http://www.arXiv.org/abs/1412.7420}{\texttt{arXiv:1412.7420}}.

\bibitem{Sjostrand:2014zea}
T.~Sj{\"o}strand\hrefCMSnoop {}{ {et~al.}, ``{An introduction to PYTHIA
  8.2}'',} \textit{ Comput. Phys. Commun.} \textbf{ 191} (2015) 159,
  \href{http://dx.doi.org/10.1016/j.cpc.2015.01.024}{\doi{10.1016/j.cpc.2015.01.024}},
\href{http://www.arXiv.org/abs/1410.3012}{\texttt{arXiv:1410.3012}}.

\bibitem{Skands:2014pea}
\hrefCMSnoop {}{P.~Skands, S.~Carrazza, and J.~Rojo, ``{Tuning PYTHIA 8.1: the
  Monash 2013 tune}'',} \textit{ Eur. Phys. J. C} \textbf{ 74} (2014) 3024,
  \href{http://dx.doi.org/10.1140/epjc/s10052-014-3024-y}{\doi{10.1140/epjc/s10052-014-3024-y}},
\href{http://www.arXiv.org/abs/1404.5630}{\texttt{arXiv:1404.5630}}.

\bibitem{CMS-PAS-GEN-14-001}
\hrefCMSnoop {}{{CMS Collaboration}, ``{Event generator tunes obtained from
  underlying event and multiparton scattering measurements}'',} \textit{ Eur.
  Phys. J. C} \textbf{ 76} (2016) 155,
  \href{http://dx.doi.org/10.1140/epjc/s10052-016-3988-x}{\doi{10.1140/epjc/s10052-016-3988-x}},
\href{http://www.arXiv.org/abs/1512.00815}{\texttt{arXiv:1512.00815}}.

\bibitem{CP5PAS}
\href {https://cds.cern.ch/record/2634082}{{CMS Collaboration}, ``{Extraction
  and validation of a new set of CMS PYTHIA 8 tunes from underlying event
  measurements}'',} CMS Physics Analysis Summary CMS-PAS-GEN-17-001, 2018.

\bibitem{Frederix:2012ps}
\hrefCMSnoop {}{R.~Frederix and S.~Frixione, ``{Merging meets matching in
  MC@NLO}'',} \textit{ JHEP} \textbf{ 12} (2012) 061,
  \href{http://dx.doi.org/10.1007/JHEP12(2012)061}{\doi{10.1007/JHEP12(2012)061}},
\href{http://www.arXiv.org/abs/1209.6215}{\texttt{arXiv:1209.6215}}.

\bibitem{Alwall:2007fs}
J.~Alwall\hrefCMSnoop {}{ {et~al.}, ``{Comparative study of various algorithms
  for the merging of parton showers and matrix elements in hadronic
  collisions}'',} \textit{ Eur. Phys. J. C} \textbf{ 53} (2008) 473,
  \href{http://dx.doi.org/10.1140/epjc/s10052-007-0490-5}{\doi{10.1140/epjc/s10052-007-0490-5}},
\href{http://www.arXiv.org/abs/0706.2569}{\texttt{arXiv:0706.2569}}.

\bibitem{Geant}
\hrefCMSnoop {}{{GEANT4} Collaboration, ``{GEANT4---a simulation toolkit}'',}
  \textit{ Nucl. Instrum. Meth. A} \textbf{ 506} (2003) 250,
\href{http://dx.doi.org/10.1016/S0168-9002(03)01368-8}{\doi{10.1016/S0168-9002(03)01368-8}}.

\bibitem{Sirunyan:2017ulk}
\hrefCMSnoop {}{{CMS Collaboration}, ``{Particle-flow reconstruction and global
  event description with the CMS detector}'',} \textit{ JINST} \textbf{ 12}
  (2017) P10003,
  \href{http://dx.doi.org/10.1088/1748-0221/12/10/P10003}{\doi{10.1088/1748-0221/12/10/P10003}},
\href{http://www.arXiv.org/abs/1706.04965}{\texttt{arXiv:1706.04965}}.

\bibitem{Cacciari:2008gp}
\hrefCMSnoop {}{M.~Cacciari, G.~P. Salam, and G.~Soyez, ``{The anti-\kt jet
  clustering algorithm}'',} \textit{ JHEP} \textbf{ 04} (2008) 063,
  \href{http://dx.doi.org/10.1088/1126-6708/2008/04/063}{\doi{10.1088/1126-6708/2008/04/063}},
\href{http://www.arXiv.org/abs/0802.1189}{\texttt{arXiv:0802.1189}}.

\bibitem{Cacciari:fastjet1}
\hrefCMSnoop {}{M.~Cacciari, G.~P. Salam, and G.~Soyez, ``{FastJet user
  manual}'',} \textit{ Eur. Phys. J. C} \textbf{ 72} (2012) 1896,
  \href{http://dx.doi.org/10.1140/epjc/s10052-012-1896-2}{\doi{10.1140/epjc/s10052-012-1896-2}},
\href{http://www.arXiv.org/abs/1111.6097}{\texttt{arXiv:1111.6097}}.

\bibitem{Cacciari:fastjet2}
\hrefCMSnoop {}{M.~Cacciari and G.~P. Salam, ``{Dispelling the $N^{3}$ myth for
  the $k_{t}$ jet-finder}'',} \textit{ Phys. Lett. B} \textbf{ 641} (2006) 57,
  \href{http://dx.doi.org/10.1016/j.physletb.2006.08.037}{\doi{10.1016/j.physletb.2006.08.037}},
\href{http://www.arXiv.org/abs/hep-ph/0512210}{\texttt{arXiv:hep-ph/0512210}}.

\bibitem{CMS-PAS-JME-10-003}
\href {http://cdsweb.cern.ch/record/1279362}{{CMS Collaboration}, ``Jet
  performance in pp collisions at $\sqrt{s} = 7$ {TeV}'',} CMS Physics Analysis
  Summary CMS-PAS-JME-10-003, 2010.

\bibitem{Chatrchyan:2011ds}
\hrefCMSnoop {}{{CMS Collaboration}, ``{Determination of jet energy calibration
  and transverse momentum resolution in CMS}'',} \textit{ JINST} \textbf{ 6}
  (2011) P11002,
  \href{http://dx.doi.org/10.1088/1748-0221/6/11/P11002}{\doi{10.1088/1748-0221/6/11/P11002}},
\href{http://www.arXiv.org/abs/1107.4277}{\texttt{arXiv:1107.4277}}.

\bibitem{Khachatryan:2016kdb}
\hrefCMSnoop {}{{CMS Collaboration}, ``{Jet energy scale and resolution in the
  CMS experiment in pp collisions at 8 TeV}'',} \textit{ JINST} \textbf{ 12}
  (2017) P02014,
  \href{http://dx.doi.org/10.1088/1748-0221/12/02/P02014}{\doi{10.1088/1748-0221/12/02/P02014}},
\href{http://www.arXiv.org/abs/1607.03663}{\texttt{arXiv:1607.03663}}.

\bibitem{Sirunyan:2017ezt}
\hrefCMSnoop {}{{CMS Collaboration}, ``{Identification of heavy-flavour jets
  with the CMS detector in pp collisions at 13 TeV}'',} \textit{ JINST}
  \textbf{ 13} (2018) P05011,
  \href{http://dx.doi.org/10.1088/1748-0221/13/05/P05011}{\doi{10.1088/1748-0221/13/05/P05011}},
\href{http://www.arXiv.org/abs/1712.07158}{\texttt{arXiv:1712.07158}}.

\bibitem{CMS-PAS-JME-16-003}
\href {https://cds.cern.ch/record/2256875}{{CMS Collaboration}, ``{Jet
  algorithms performance in 13 TeV data}'',} CMS Physics Analysis Summary
  CMS-PAS-JME-16-003, 2016.

\bibitem{Khachatryan:2015hwa}
\hrefCMSnoop {}{{CMS Collaboration}, ``{Performance of electron reconstruction
  and selection with the CMS detector in proton-proton collisions at $\sqrt{s}
  = 8\TeV$}'',} \textit{ JINST} \textbf{ 10} (2015) P06005,
  \href{http://dx.doi.org/10.1088/1748-0221/10/06/P06005}{\doi{10.1088/1748-0221/10/06/P06005}},
\href{http://www.arXiv.org/abs/1502.02701}{\texttt{arXiv:1502.02701}}.

\bibitem{Sirunyan:2018fpa}
\hrefCMSnoop {}{{CMS Collaboration}, ``{Performance of the CMS muon detector
  and muon reconstruction with proton-proton collisions at $\sqrt{s}=$ 13
  TeV}'',} \textit{ JINST} \textbf{ 13} (2018) P06015,
  \href{http://dx.doi.org/10.1088/1748-0221/13/06/P06015}{\doi{10.1088/1748-0221/13/06/P06015}},
\href{http://www.arXiv.org/abs/1804.04528}{\texttt{arXiv:1804.04528}}.

\bibitem{Khachatryan:2016kod}
\hrefCMSnoop {}{{CMS Collaboration}, ``{Search for new physics in same-sign
  dilepton events in proton-proton collisions at $\sqrt{s} = 13$ TeV}'',}
  \textit{ Eur. Phys. J. C} \textbf{ 76} (2016) 439,
  \href{http://dx.doi.org/10.1140/epjc/s10052-016-4261-z}{\doi{10.1140/epjc/s10052-016-4261-z}},
\href{http://www.arXiv.org/abs/1605.03171}{\texttt{arXiv:1605.03171}}.

\bibitem{Hocker:2007ht}
\hrefCMSnoop {}{H.~Voss, A.~H{\"o}cker, J.~Stelzer, and F.~Tegenfeldt,
  ``{TMVA}, the toolkit for multivariate data analysis with {ROOT}'',} in
  \textit{ XIth International Workshop on Advanced Computing and Analysis
  Techniques in Physics Research (ACAT)}, p.~40.
\newblock 2007.
\newblock
  \href{http://www.arXiv.org/abs/physics/0703039}{\texttt{arXiv:physics/0703039}}.
\newblock {[PoS(ACAT)040]}.
\href{http://dx.doi.org/10.22323/1.050.0040}{\doi{10.22323/1.050.0040}}.

\bibitem{chollet2015keras}
\hrefCMSnoop {}{F.~Chollet {et~al.}, ``Keras'',} 2015.
\newblock Software available from \url{https://keras.io}.

\bibitem{tensorflow2015-whitepaper}
\hrefCMSnoop {}{M.~Abadi {et~al.}, ``{TensorFlow}: Large-scale machine learning
  on heterogeneous systems'',} (2016).
  \href{http://www.arXiv.org/abs/1603.04467}{\texttt{arXiv:1603.04467}}.
  Software available from \url{https://www.tensorflow.org/}.

\bibitem{PDG}
\hrefCMSnoop {}{{Particle Data Group}, M.~Tanabashi {et~al.}, ``Review of
  particle physics'',} \textit{ Phys. Rev. D} \textbf{ 98} (2018) 030001,
  \href{http://dx.doi.org/10.1103/PhysRevD.98.030001}{\doi{10.1103/PhysRevD.98.030001}}.

\bibitem{Sirunyan:2018nqx}
\hrefCMSnoop {}{{CMS Collaboration}, ``{Measurement of the inelastic
  proton-proton cross section at $ \sqrt{s}=13 $ TeV}'',} \textit{ JHEP}
  \textbf{ 07} (2018) 161,
  \href{http://dx.doi.org/10.1007/JHEP07(2018)161}{\doi{10.1007/JHEP07(2018)161}},
\href{http://www.arXiv.org/abs/1802.02613}{\texttt{arXiv:1802.02613}}.

\bibitem{CMS-PAS-LUM-17-001}
\href {http://cds.cern.ch/record/2257069}{{CMS Collaboration}, ``{CMS}
  luminosity measurements for the 2016 data-taking period'',} CMS Physics
  Analysis Summary CMS-PAS-LUM-17-001, 2017.

\bibitem{CMS-PAS-LUM-17-004}
\href {http://cds.cern.ch/record/2621960}{{CMS Collaboration}, ``{CMS}
  luminosity measurement for the 2017 data-taking period at 13 {TeV}'',} CMS
  Physics Analysis Summary CMS-PAS-LUM-17-004, 2018.

\bibitem{Butterworth:2015oua}
\hrefCMSnoop {}{J.~Butterworth {et~al.}, ``{PDF4LHC recommendations for LHC Run
  II}'',} \textit{ J. Phys. G} \textbf{ 43} (2016) 023001,
  \href{http://dx.doi.org/10.1088/0954-3899/43/2/023001}{\doi{10.1088/0954-3899/43/2/023001}},
\href{http://www.arXiv.org/abs/1510.03865}{\texttt{arXiv:1510.03865}}.

\bibitem{Khachatryan:2014jba}
\hrefCMSnoop {}{{CMS Collaboration}, ``{Precise determination of the mass of
  the Higgs boson and tests of compatibility of its couplings with the standard
  model predictions using proton collisions at 7 and 8 $\,\text {TeV}$}'',}
  \textit{ Eur. Phys. J. C} \textbf{ 75} (2015) 212,
  \href{http://dx.doi.org/10.1140/epjc/s10052-015-3351-7}{\doi{10.1140/epjc/s10052-015-3351-7}},
\href{http://www.arXiv.org/abs/1412.8662}{\texttt{arXiv:1412.8662}}.

\bibitem{Cowan:2010js}
\hrefCMSnoop {}{G.~Cowan, K.~Cranmer, E.~Gross, and O.~Vitells, ``Asymptotic
  formulae for likelihood-based tests of new physics'',} \textit{ Eur. Phys. J.
  C} \textbf{ 71} (2011) 1554,
  \href{http://dx.doi.org/10.1140/epjc/s10052-011-1554-0}{\doi{10.1140/epjc/s10052-011-1554-0}},
  \href{http://www.arXiv.org/abs/1007.1727}{\texttt{arXiv:1007.1727}}.
[Erratum: \DOI{10.1140/epjc/s10052-013-2501-z}].

\bibitem{ATLAS:1379837}
\href {https://cds.cern.ch/record/1379837}{{The ATLAS Collaboration, The CMS
  Collaboration, The LHC Higgs Combination Group}, ``Procedure for the {LHC}
  {Higgs} boson search combination in {Summer} 2011'',} Technical Report
  CMS-NOTE-2011-005, ATL-PHYS-PUB-2011-11, 2011.

\end{thebibliography}\endgroup
\ifthenelse{\boolean{cms@external}}{}{
\clearpage
\appendix
\numberwithin{table}{section}
\numberwithin{figure}{section}
\section{Supplemental information\label{app:suppMat}}
\begin{table*}[tbh]
\centering
\topcaption{Post-fit expected background and \tZq signal event yields with their total uncertainties, and observed number of events in data in each of the signal regions.}
\label{tab:yields}
\cmsTable{
\begin{scotch}{cccccc}
Source & 2--3 jets, 1 \PQb-tagged & $\geq$ 4 jets, 1 \PQb-tagged & $\geq$ 2 \PQb jets & \WZ enriched & \ZZ enriched\\
       & SR-2/3j-1b & SR-4j-1b & SR-2b & & \\ \hline
Exp. background & $357 \pm 34$ & $278 \pm 32$ & $228 \pm 25$ & $6308 \pm 478$ & $ 847 \pm 69$\\
Exp. \tZq & $103 \pm 5.1$ & $38 \pm 5.3$ & $37 \pm 1.8$ & $67 \pm 2.8$ & $0.01 \pm 0.004 $\\
Total exp. & $460 \pm 37$ & $316 \pm 35$ & $265 \pm 25$ & $6375 \pm 478$ & $847 \pm 69$ \\
Observed &$475$ & $310$ & $278$ & $6373$ & $852$ \\
\end{scotch}}
\end{table*}

\begin{figure*}[htbp]
\includegraphics[width=.325\textwidth]{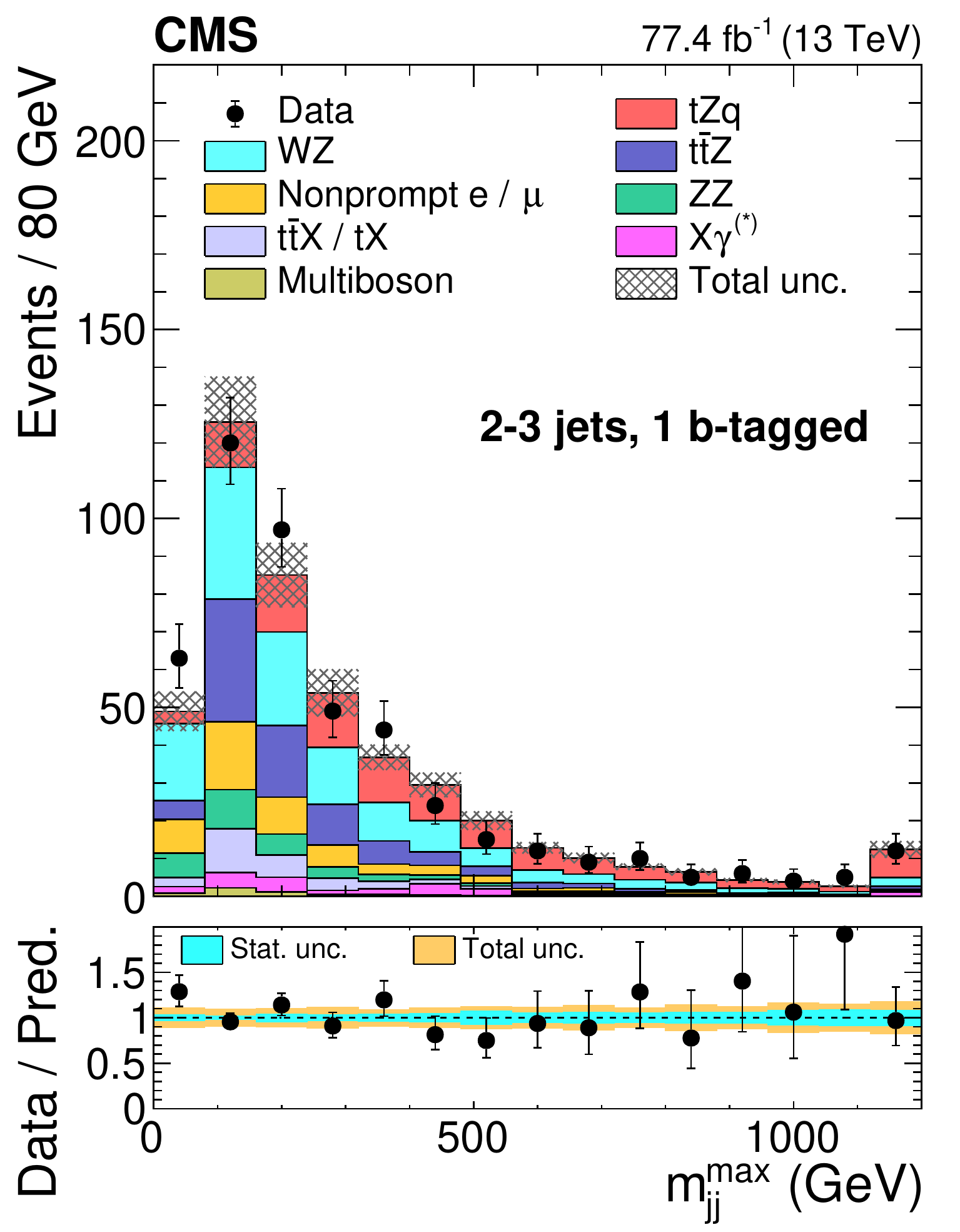}
\includegraphics[width=.325\textwidth]{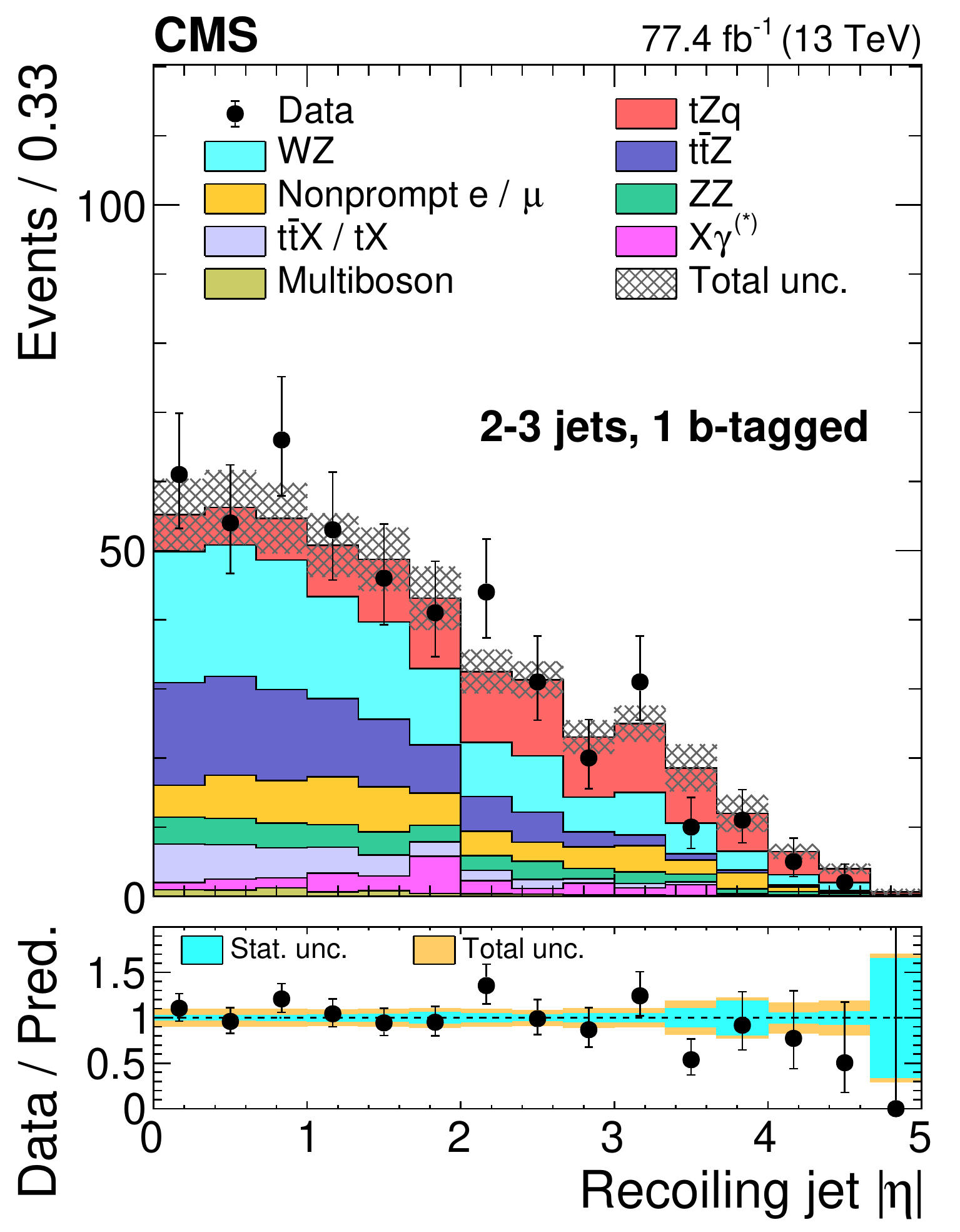}
\includegraphics[width=.325\textwidth]{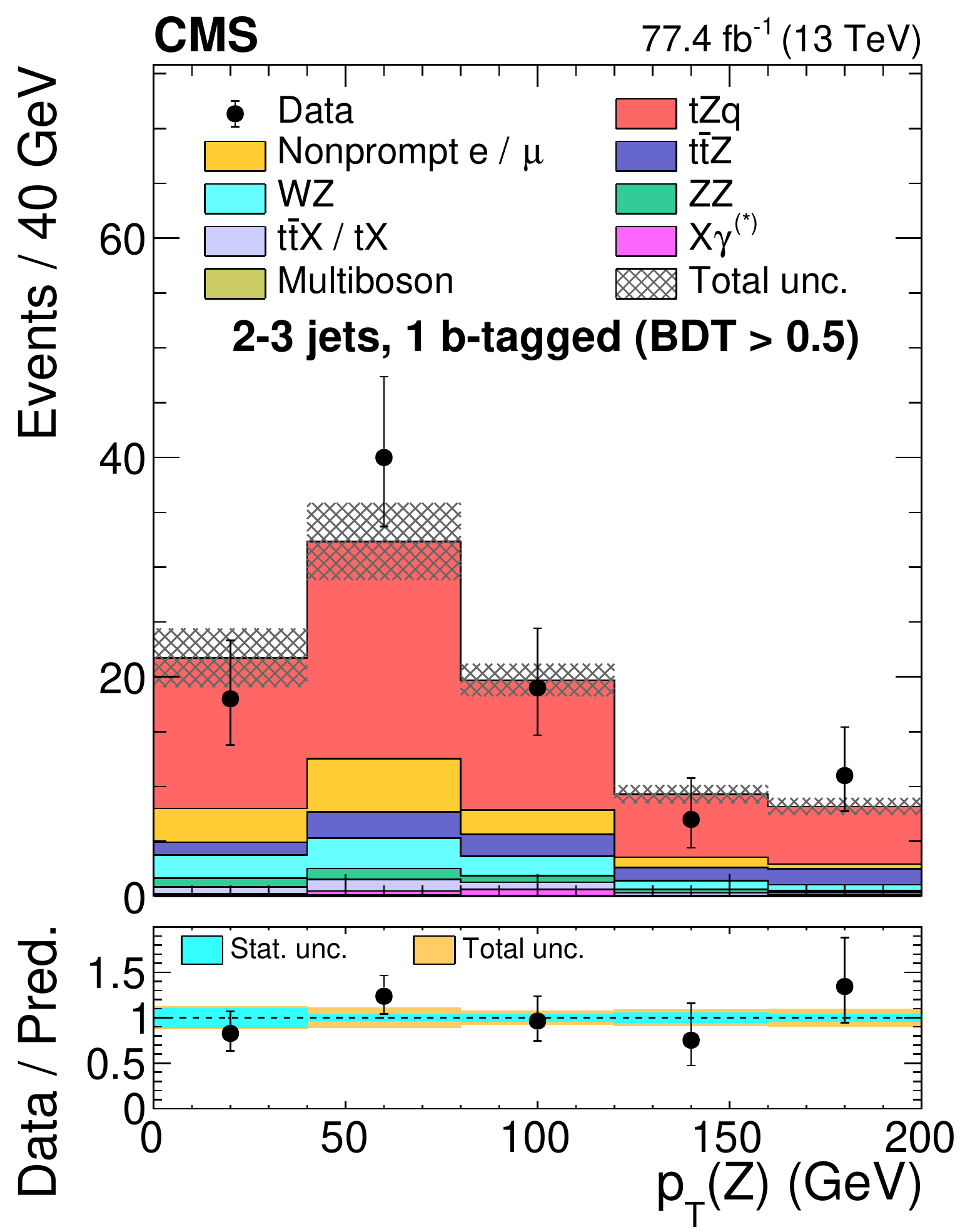}
\caption{Observed (points) and post-fit expected distributions (shaded histograms) in SR-2/3j-1b events for the two most discriminating variables used in the BDT discriminant, the maximum dijet invariant mass among all pairs of jets in the event (left), and the \abseta of the recoiling jet (middle). The right plot shows the \pt of the \PZ boson, reconstructed from its leptonic decay products, for events with BDT discriminant values in excess of 0.5 in SR-2/3j-1b. This observable is highly sensitive to the presence of new physics phenomena. The vertical bars on the points give the statistical uncertainty in data, and the hatched regions display the total uncertainty in the prediction. The lower panels display the ratio of the observed data to the predictions, including the \tZq signal, with inner and outer shaded bands, respectively, representing the statistical and total uncertainties in the predictions.}
\label{fig:supplementaryplots}
\end{figure*}

\begin{figure*}[htbp]
\centering
\includegraphics[width=.325\textwidth]{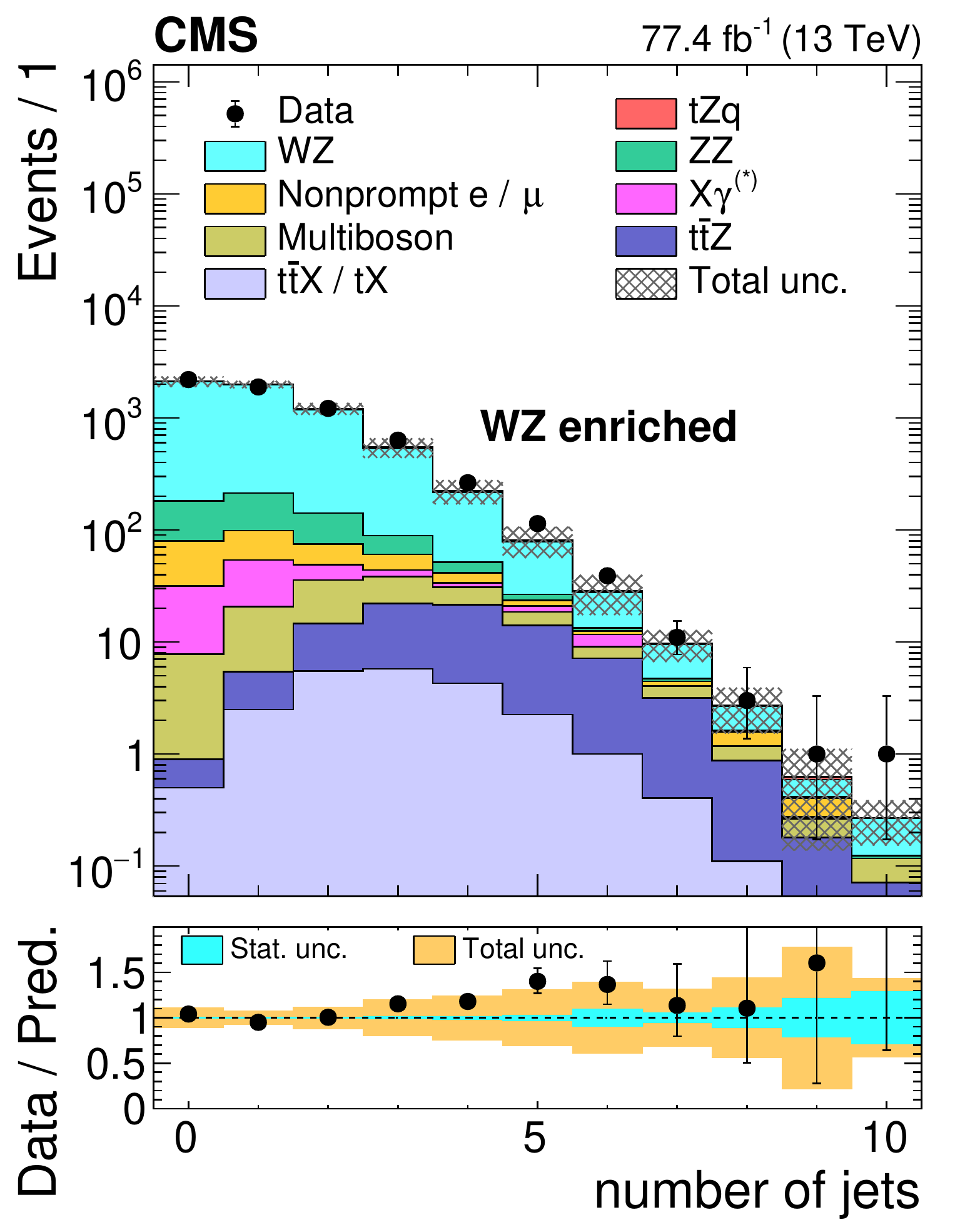}
\includegraphics[width=.325\textwidth]{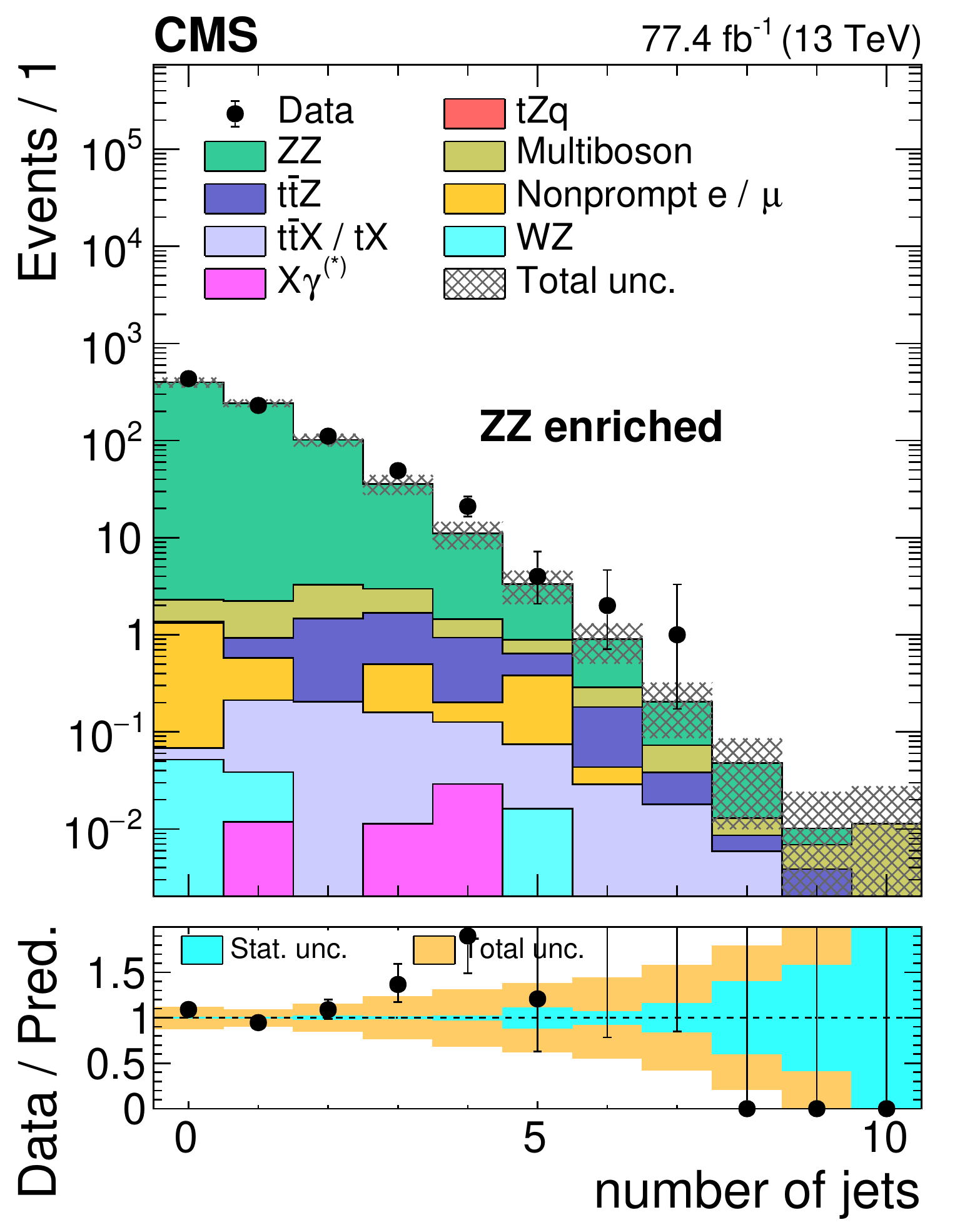}
\caption{Observed (points) and pre-fit expected distributions (shaded histograms) of the number of jets in the event for the \WZ (left), and \ZZ (right) control regions. The vertical bars on the points give the statistical uncertainty in data, and the hatched regions display the total uncertainty in the prediction. The lower panels display the ratio of the observed data to the predictions, including the \tZq signal, with inner and outer shaded bands, respectively, representing the statistical and total uncertainties in the predictions.}
\end{figure*}

{\begin{table}[tbh]
\centering
\caption{Average impact on the measured \tZq signal strength for major sources of systematic uncertainty. The impact of a particular nuisance parameter on the signal strength is computed by shifting the nuisance parameter by one standard deviation up and down from its post-fit value and recomputing the signal strength. All other nuisances are profiled as in the nominal fit when doing this computation.}
\label{tab:systematics_impacts}
\centering
\begin{scotch}{lr}
Uncertainty & Impact (\%) \\ 
\hline
\multicolumn{2}{c}{ Experimental } \\ 
\hline
lepton selection & 3.2 \\
trigger efficiency & 1.4 \\
jet energy scale & 3.3 \\
b-tagging efficiency & 1.7 \\
nonprompt normalization & 4.1\\
\ttZ normalization & 1.0 \\
luminosity & 1.7\\
pileup & 1.9\\
other& 1.3\\
\multicolumn{2}{c}{ Theoretical } \\ 
\hline
final-state radiation & 2.0\\
tZq QCD scale & 2.0 \\
\ttZ QCD scale & 1.4 \\
\end{scotch}
\end{table}}

}
\cleardoublepage \section{The CMS Collaboration \label{app:collab}}\begin{sloppypar}\hyphenpenalty=5000\widowpenalty=500\clubpenalty=5000\vskip\cmsinstskip
\textbf{Yerevan Physics Institute, Yerevan, Armenia}\\*[0pt]
A.M.~Sirunyan, A.~Tumasyan
\vskip\cmsinstskip
\textbf{Institut f\"{u}r Hochenergiephysik, Wien, Austria}\\*[0pt]
W.~Adam, F.~Ambrogi, E.~Asilar, T.~Bergauer, J.~Brandstetter, M.~Dragicevic, J.~Er\"{o}, A.~Escalante~Del~Valle, M.~Flechl, R.~Fr\"{u}hwirth\cmsAuthorMark{1}, V.M.~Ghete, J.~Hrubec, M.~Jeitler\cmsAuthorMark{1}, N.~Krammer, I.~Kr\"{a}tschmer, D.~Liko, T.~Madlener, I.~Mikulec, N.~Rad, H.~Rohringer, J.~Schieck\cmsAuthorMark{1}, R.~Sch\"{o}fbeck, M.~Spanring, D.~Spitzbart, W.~Waltenberger, J.~Wittmann, C.-E.~Wulz\cmsAuthorMark{1}, M.~Zarucki
% ...
\vskip\cmsinstskip
% ...
1:  Also at Vienna University of Technology, Vienna, Austria\\
% ...
\end{sloppypar}
\end{document}